\documentclass{article}

\usepackage[intlimits]{amsmath}
\usepackage{cite}
\usepackage{epsfig}

\def\bea{\begin{eqnarray}}
\def\eea{\end{eqnarray}}

\def\dash{\,\textendash\, }
\def\lmatrix{\left(\begin{array}}
\def\rmatrix{\end{array}\right)}

\begin{document}

\vspace{-1cm}

\hfill UCSD/PTH 09-04

\vspace{1cm}

\begin{center}{\Large\bf Topology and higher dimensional representations}\\

\vspace{2cm}

Zolt\'an Fodor$^{\,abcd}$,$\;$ Kieran Holland$^{\,e}$,$\;$ Julius Kuti$^{\,f}$,\\

\vspace{0.2cm}

D\'aniel N\'ogr\'adi$^{\,f}\,$ and$\,$ Chris Schroeder$^{\,f}$

\vspace{1cm}

\end{center}

\hspace{-0.1cm} $^a${\em University of Wuppertal, Department of Theoretical Physics, Wuppertal 42119, Germany}

\vspace{0.2cm}

$^b${\em NIC/DESY Zeuthen Forschungsgruppe, D-15738 Zeuthen, Germany }

\vspace{0.2cm}

$^c${\em Forschungszentrum J\"ulich, D-52425 J\"ulich, Germany }

\vspace{0.2cm}

$^d${\em E\"otv\"os University, Institute for Theoretical Physics, Budapest 1117, Hungary }

\vspace{0.2cm}

$^e${\em University of the Pacific, Department of Physics, Stockton, CA 95211, USA}

\vspace{0.2cm}

$^f${\em University of California, San Diego, Department of Physics, La Jolla, CA 92093, USA }

\vspace{2cm}

\begin{abstract}
$SU(3)$ gauge theory in the 2-index symmetric (sextet) and fundamental representations is considered in symmetric and periodic
boxes. Using the overlap formulation in the quenched
approximation it is shown that the topological charge obtained from the sextet index theorem always leads
to an integer value and agrees with the charge obtained from the fundamental index theorem in the continuum.
At larger lattice spacing configurations exist with fractional
topological charge if the sextet index is used but these are lattice artifacts and the probability of finding such a configuration
rapidly approaches zero. By considering the
decomposition of the sextet representation with respect to an $SU(2)$ subgroup it is shown that the $SU(2)$ adjoint index theorem leads
to integer charge as well. We conclude
that the non-zero value of the bilinear gaugino condensate in ${\cal N}=1$ super-Yang-Mills theory cannot be attributed to
configurations with fractional topological charge once periodic boundary conditions are imposed.
\end{abstract}

\newpage

\section{Introduction}

The chirally invariant overlap formulation is known to have advantageous topological and continuum properties in the fundamental representation
\cite{Hasenfratz:1998ri, Niedermayer:1998bi, Adams:1998eg, Adams:1999if, Adams:2000rn},
but much less is known in higher dimensional representations. A naive use of the Atiyah-Singer index theorem at finite
lattice spacing may lead to fractional topological charge $Q = I / 2T$ where $T$ is the trace normalization factor of the
representation and $I$ is the index of the Dirac operator, if $I$ is not an integer multiple of $2T$. 

Here we study the 2-index symmetric (sextet) representation of $SU(3)$ in the quenched approximation. In this case $2T=5$.
This model has been considered previously in \cite{Kogut:1984sb, Shamir:2008pb, Svetitsky:2008bw, DeGrand:2008dh, Fodor:2008hm, DeGrand:2008kx}.

Our motivations for this work are twofold. 
First, the theory of $N_f = 2$ fermions in the sextet of $SU(3)$ is a promising candidate for a walking technicolor model \cite{Dietrich:2006cm}. 
The $N_f = 0$
theory certainly breaks chiral symmetry and is a good testing ground for chiral and/or topological properties of sextet overlap
fermions before addressing dynamical questions such as the infrared properties of the $N_f = 2$ theory. Results on non-abelian
gauge theories with fermions in higher dimensional representations have been reported in 
\cite{Edwards:1998dj, Damgaard:2001fg, DelDebbio:2008zf, Catterall:2008qk, Hietanen:2008vc, DelDebbio:2008tv, Hietanen:2008mr, Hietanen:2009az}.

Second, in $SU(N)$ ${\cal N} = 1$ super-Yang-Mills theory the gaugino condensate involving $2N$ gaugino fields is non-zero due to the
contribution of instantons to this correlation function since there are $2N$ adjoint zero modes for a
charge $1$ instanton. The general expectation is that the 
bilinear gaugino condensate is non-zero as well \cite{Veneziano:1982ah}. However it is not clear if this non-zero value can be attributed to some special class of
configurations if periodic boundary conditions are used. Since the fermionic definition of the topological charge can lead to
fractional values at finite lattice spacing as mentioned above, it is worth investigating what happens to these configurations in
the continuum limit. If the fractional value $1/N$ does survive the continuum limit with periodic boundary conditions
they potentially can explain the non-zero value of the bilinear 
gaugino condensate on $R^4$. This would then be similar to the case of $S^1 \times R^3$ where fractional topological charge
configurations do exist and it has been argued that these can explain the gaugino condensate \cite{Cohen:1983fd}.

The main result of our paper is that once periodic boundary conditions are imposed, configurations where $Q = I/2T$ is not an
integer disappear in the continuum limit. In the continuum all $SU(3)$
configurations respect the sextet index theorem, i.e. zero modes come in 5-tuples, furthermore the topological charge defined
by the sextet index agrees with the one from the fundamental index and is an integer.

Although we have only measured the sextet and fundamental index for $SU(3)$ with periodic boundary conditions 
the following property of continuum gauge fields
provides an opportunity to address the supersymmetric case of $SU(2)$ as well. An $SU(N)$ gauge field can
always be deformed to lie in an $SU(2)$ subgroup and the representation then can be decomposed with respect to this subgroup. In
our case the decomposition of the sextet of $SU(3)$ contains both the fundamental and adjoint of $SU(2)$. This observation leads to the conclusion that
in $SU(2)$ gauge theory the fundamental and adjoint index theorems both give the same definition of the topological charge in the
continuum. In particular, the charge is an integer. 

Consequently, the bilinear gaugino condensate
cannot be attributed to fractionally charged configurations subject to periodic boundary conditions.

Although we have not investigated other higher dimensional representations, we conjecture that our findings hold
more generally. Specifically, in a large periodic box,
(1) the fermionic definition of the topological charge $Q = I / 2T $ will always be an integer and (2) two definitions from two
different representations will agree in the continuum limit.

The adjoint representation of $SU(2)$ was investigated directly in \cite{Edwards:1998dj} and a slight increase has been reported
in the probability of finding a configuration with adjoint index $2$, suggesting $Q=1/2$, as the lattice spacing decreased at
fixed physical volume.
This raised the possibility that the probability stays finite in the continuum limit.
We will show however that if the lattice spacing is too large the decrease in this probability with decreasing lattice spacing is
quite small and can easily be confused with a slight increase at low statistics. A
more consistent decrease of the probability with decreasing lattice spacing does however set in for sufficiently small lattice
spacing. 

\section{Topology and index theorem}

The Atiyah-Singer index theorem in the continuum states that $I_R = 2T_R Q$ where $I_R$ is the index of the Dirac operator in representation
$R$, $T_R$ is its trace normalization factor and $Q$ is the topological charge of the underlying gauge field. At finite lattice spacing a
gluonic definition of $Q$ is not available in general but the above relation can be employed to define it using the overlap
operator which possesses exact chiral zero modes \cite{Neuberger:1997fp}. Clearly, two different representations may lead to different charges for
the same configuration. Also, if $2T_R > 1$ the fermionic definition may lead to non-integer charge. In the fundamental
representation this problem does not arise because $2T = 1$ and the fermionic definition always gives an integer charge (different choices for
the negative Wilson mass may still lead to different integers but in the continuum limit the charge will not be sensitive to this
choice).

For gauge group $SU(N)$ the $T$ factors for some of the most frequently used representations are given by,
\bea
\label{T}
T_F = \frac{1}{2}\;,\quad T_{adj} = N\;,\quad T_S = \frac{N+2}{2}\;,\quad T_A = \frac{N-2}{2}\;,
\eea
where $F$, $adj$, $S$ and $A$ refer to the fundamental, adjoint, 2-index symmetric and 2-index anti-symmetric representations, respectively.
The index for $SU(N)$ will be denoted by $I^N_R$. Unless otherwise indicated $Q$ and $I$ will stand for the absolute value of these
quantities.

\subsection{Sextet representation of SU(3)}

For the sextet of $SU(3)$ we have $2T = 5$ thus the continuum index theorem suggests that zero modes of the Dirac
operator ought to come in 5-tuples.

We have measured the index of the Dirac operator in both the fundamental and sextet representation in the quenched approximation
at 5 different lattice spacings at the same physical volume. For a similar study in the fundamental representation only, see \cite{Giusti:2003gf}.
The index was obtained by calculating the first few eigenvalues of the Dirac
operator until a non-zero eigenvalue is found and checking whether they are chiral or not. 
Zero modes are clearly identified by their chirality being $\pm1$ whereas
non-zero modes have 0 chirality just as in the continuum. 
The non-zero eigenvalues carry important chiral information encoded in the
Banks-Casher relation and random matrix theory. These findings will be reported elsewhere \cite{future}. 

The Wilson plaquette action was used on symmetrical $L^4$ lattices
with periodic boundary conditions and the Sommer scale $r_0 = 0.5\,fm$ (in the
fundamental representation) was used to set the scale \cite{Necco:2001xg}. The negative Wilson mass in the fundamental representation was chosen at
$-1.4$ while for the sextet it was set to $-1.7$. In both cases 2 steps of stout smearing was applied with smearing parameter
$0.15$ \cite{Morningstar:2003gk}. The implementation of the overlap operator is the same as in \cite{Fodor:2003bh}.
The simulation parameters are summarized in Table \ref{simparam}. The fundamental and sextet
indices are tabulated in Tables \ref{indices12} \dash \ref{indices20} together with the number of configurations that have been
observed with the given values.

\begin{table}
\begin{center}
\begin{tabular}{|ccccc|}
\hline
$\beta$ & $L/a$ & $r_0/a$ & $a\;[fm]$ & \#config \\
\hline
\hline
5.9500  & 12  & 4.9122 & 0.1018 & 441 \\
\hline
6.0384  & 14  & 5.7306 & 0.0873 & 543 \\
\hline
6.1210  & 16  & 6.5496 & 0.0763 & 408 \\
\hline
6.2719  & 20  & 8.1867 & 0.0611 & 211 \\
\hline
6.4064  & 24  & 9.8239 & 0.0509 & 105 \\
\hline
\end{tabular}
\caption{Simulation parameters using the Wilson plaquette gauge action. The physical volume is kept fixed at $L=1.2215\,fm$.}
\label{simparam}
\end{center}
\end{table}

\begin{figure}
\begin{center}
\begin{tabular}{cc}
\includegraphics[width=6.5cm]{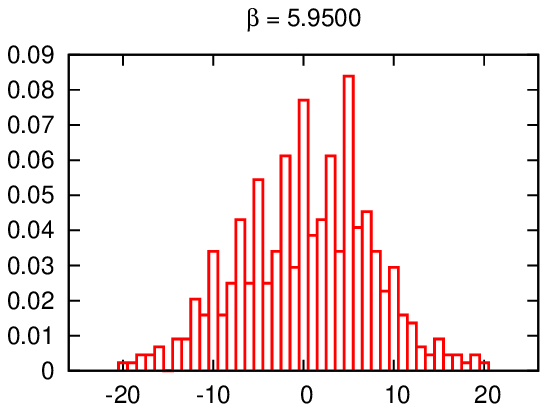} & \includegraphics[width=6.5cm]{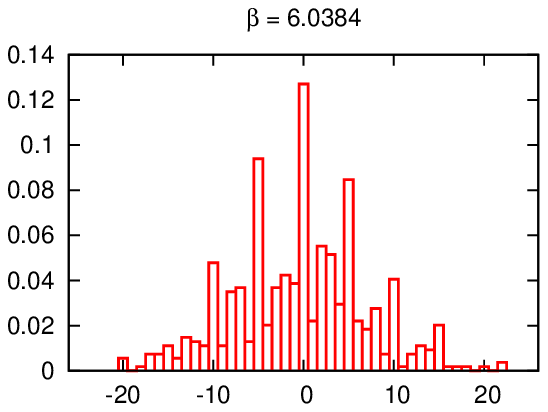} \\
\includegraphics[width=6.5cm]{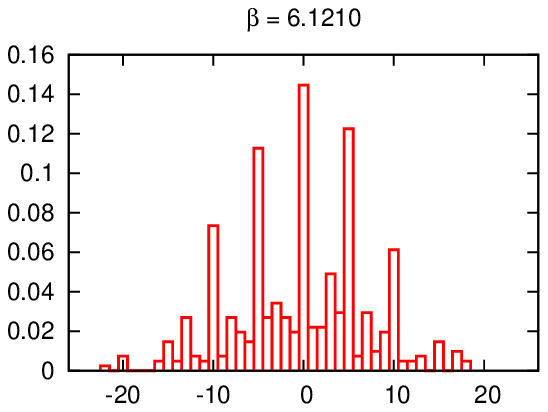} & \includegraphics[width=6.5cm]{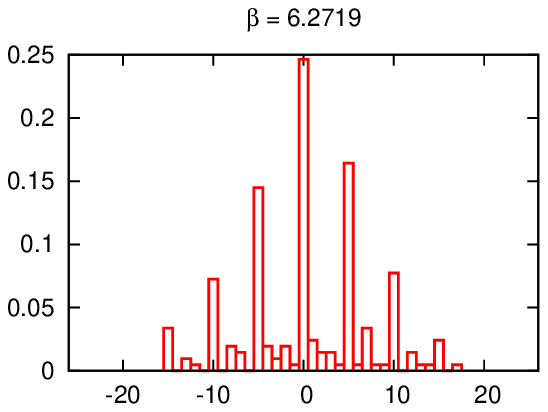} \\
\includegraphics[width=6.5cm]{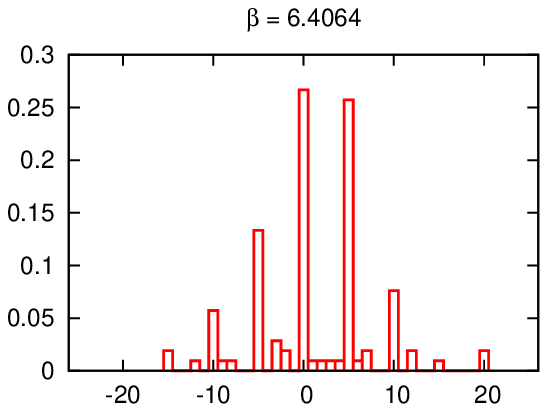} & 
\end{tabular}
\caption{Normalized distribution of the $SU(3)$ sextet index at fixed physical volume and various lattice spacings.\label{zmdistr}}
\end{center}
\end{figure}

\begin{figure}
\begin{center}
\begin{tabular}{cc}
\includegraphics[width=11cm]{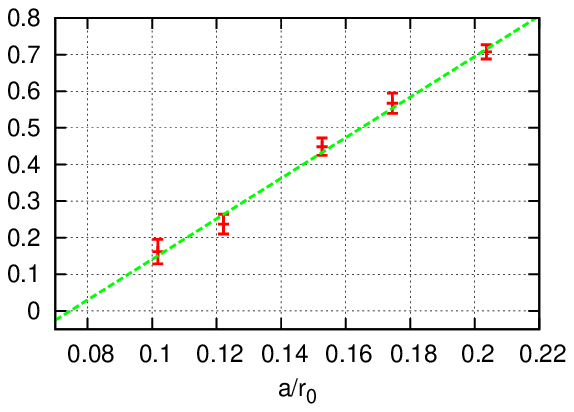}
\end{tabular}
\caption{Probability of finding a configuration where the number of sextet zero modes is not a multiple of $5$.\label{zmlimit}}
\end{center}
\end{figure}

First, let us analyze to what extent sextet zero modes come in 5-tuples. 
The distribution of the 
sextet index is shown on Figure \ref{zmdistr} without taking the absolute value. Clearly, as the lattice spacing decreases the number of configurations that
would lead to fractional charge is suppressed.
The probability that one finds such a configuration is shown on
Figure \ref{zmlimit}. 

Fitting this probability to a function of $a$ requires some care. Unfortunately we do not have a solid theoretical expectation for
the $a$-dependence as a whole and neither for the limit $a\to 0$. In principle the $a\to 0$ limit could be worked out analytically
and we hope to return to this question in the future. Currently, however, the only guide is the data itself. 
The chiral symmetry of overlap fermions ensure $O(a)$ improvement in
on-shell correlation functions of local operators \cite{Niedermayer:1998bi}, but the probability plotted on Figure \ref{zmlimit} is not such a quantity.
Hence as $a\to 0$ we do not expect $O(a)$ terms to be absent. 

Although the linear fit to the data does not go through zero at $a = 0$, the actual
linear fit does seem to work very well. We cannot completely rule out the
scenario that the probability of finding a configuration with sextet indices that are not multiples
of $5$ will only vanish at zero lattice spacing, deviating from the fitted straight line 
for smaller $a$ values than the simulation points. However, our
current data set is consistent with the linear fit, reaching zero at finite lattice spacing. 


Keeping in mind these cautionary remarks we proceed with the linear fit.
It implies that at a certain small but finite lattice spacing all configurations that are not consistent with the
sextet index theorem and integer charge are gone. The extrapolated
value is $a_c/r_0 = 0.075(8)$ corresponding to $a_c = 0.038(4)\,fm$. At our physical volume this corresponds to approximately
$\beta = 6.64$ and $L/a = 32$. The precise value of $a_c$ presumably
depends on the fixed finite physical volume, as well as on the details of the fermion matrix such
as the number of smearing steps and the smearing parameter. Naturally, it would be most interesting to calculate the sextet 
and fundamental indices on lattices finer than $a_c$
to verify the validity of the extrapolation used, but currently we do not have sufficient resources. 

Second, let us investigate how many configurations violate the continuum expectation $5I^3_F = I^3_S$. 
Their number is actually almost the same as the number of configurations with $I^3_S$ not a multiple of $5$, see Tables
\ref{indices12} \dash \ref{indices20}.
Correspondingly, the
extrapolated lattice spacing value $a_c$ where all configurations with the property $5I^3_F \neq I^3_S$ disappear is the same, 
within errors, as the one obtained above. Our comments about the $a\to 0$ behaviour above naturally apply to this case as well.

Summarizing this section we conclude that if $a<a_c$, or perhaps only in the continuum limit,
the only remaining configurations are the ones where $I^3_S$ is an integer multiple of
$5$ and in addition the topological charge defined
from the fundamental representation coincides with the definition from the sextet representation. In particular it is an integer.

\subsection{Adjoint representation of SU(2)}

Although we have only studied the fundamental and sextet representations of $SU(3)$ our results have implications for the
adjoint representation of $SU(N)$ as well. This is because in the continuum an $SU(N)$ gauge field can always be deformed to lie
in an $SU(2)$ subgroup only. The fermion representation can then be decomposed with respect
to this $SU(2)$ subgroup. For the sextet of $SU(3)$ this decomposition is
\bea
\label{sexdeco}
{\bf 6}_{SU(3)} = {\bf 3}_{SU(2)} \oplus {\bf 2}_{SU(2)} \oplus {\bf 1}_{SU(2)}\;,
\eea
containing the adjoint and fundamental of $SU(2)$.

From the relations (\ref{T}) we find that an $SU(2)$ charge $1$ gauge field will carry $4$ adjoint and $1$
fundamental zero modes, $I^2_{adj} = 4Q$ and $I^2_F=Q$. This also confirms that for $SU(3)$ there are $4 + 1 = 5$ sextet zero modes. 
These $5$ sextet zero modes of a charge
$1$ $SU(3)$ gauge field can be grouped into a group of $4$ which are the $4$ adjoint $SU(2)$ zero modes and a group containing just a
single fundamental $SU(2)$ zero mode.

This is the reason why our sextet $SU(3)$ zero modes know about adjoint $SU(2)$ zero modes.

Since we have seen in the previous section that the $SU(3)$ sextet index is compatible with integer charge if the lattice spacing
is small enough, it strongly suggests that the $SU(2)$ adjoint index is as well. To make this claim more quantitative,
we have $I^3_S - I^3_F = I^2_{adj}$ for an $SU(2)$ gauge field embedded in $SU(3)$ using the decomposition (\ref{sexdeco}), since
in this case $I^2_F = I^3_F$. If
the $SU(2)$ adjoint index theorem is not compatible with integer charge, $I^3_S - I^3_F$ should not be a multiple of $4$.
Inspecting the Tables \ref{indices12} \dash \ref{indices20} we find that once $I^3_S - I^3_F$ is not a multiple of $4$, the remainder is
almost always $2$, i.e. $I^3_S - I^3_F = 4k + 2$ suggesting $Q=k+1/2$. In \cite{Edwards:1998dj} the reported violation of the $SU(2)$
adjoint index theorem was also found in this channel.

\begin{figure}
\begin{center}
\begin{tabular}{cc}
\includegraphics[width=11cm]{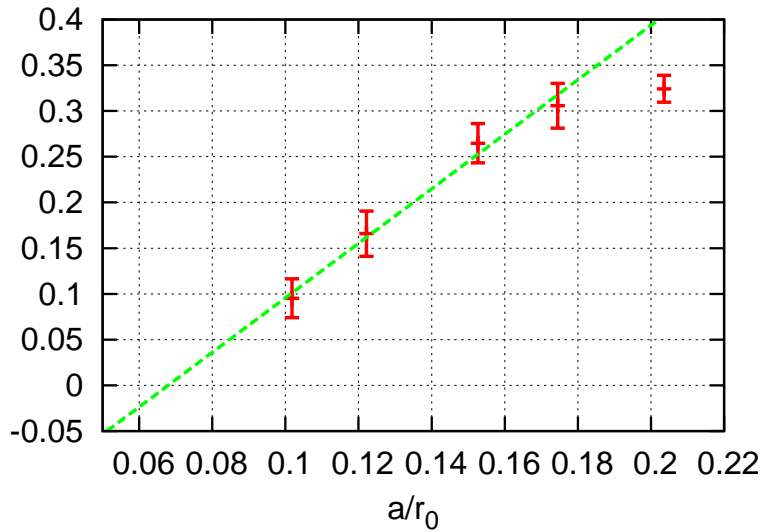}
\end{tabular}
\caption{Probability of finding configurations with $I^3_S-I^3_F = 4k + 2$ for integer $k$. These can be interpreted as embedded $SU(2)$
configurations with $I^2_{adj} = 4k + 2$ suggesting $Q=k+1/2$.\label{zmsu2limit}}
\end{center}
\end{figure}

On Figure \ref{zmsu2limit} we plot the probability of finding a configuration with the property $I^3_S - I^3_F = 4k+2$. It is
worth noting that the decrease in this probability from the largest to the second largest lattice spacing is consistent with zero.
At smaller lattice spacing a clear decreasing trend sets in which again can be fitted with a linear
$a$-dependence, omitting the largest lattice spacing. The extrapolated $a_c$ value where all
such configurations disappear is $a_c/r_0 = 0.068(12)$, compatible with the one obtained in the previous section. Naturally, all
the comments in the previous section about the $a\to 0$ limit and the fitting procedure itself apply to this case as well.

There are also other ways of measuring how the $SU(2)$ index theorem is violated. For example, configurations with $I^2_{adj} =
4k+2$ presumably have $I^2_F = k$ or $k+1$. Consequently, we have $|5I^3_F -I^3_S| = 2$ or $3$ for all $k$. The probability of
finding configurations with this property behaves very similarly to Figure \ref{zmsu2limit}. Or the probability of finding configurations with the
property $I^3_S = 5k+2$ or $5k+3$ again behaves similarly. The decrease in all of these probabilities from the largest to the
second largest lattice spacing is either consistent with zero or just barely outside $1$ standard deviation. All of these observations lead to
the conclusion that the low statistics is the most probable reason why the absence of configurations with $I^2_{adj} = 2$ was
not ruled out in the continuum limit in \cite{Edwards:1998dj}. The lattice spacings are difficult to compare since that
simulation was directly for the gauge group $SU(2)$ though.

From our analysis we see however that in the continuum the adjoint $SU(2)$ index theorem is compatible
with integer charge and configurations with fractional charge are lattice artifacts. 
In this respect the heuristic arguments in \cite{GarciaPerez:2007ne} suggesting that the configurations in
\cite{Edwards:1998dj} might have been too rough may actually be correct.

Instantons give rise to a non-zero condensate
involving $2N$ gauginos in $SU(N)$ ${\cal N}=1$ super-Yang-Mills theory because each instanton carries $2N$ zero modes. 
The general expectation is that the bilinear gaugino condensate
is non-zero as well. Non-periodic boundary conditions are capable of supporting gauge fields with fractional topological charge
and it has been argued that these configurations are responsible for the non-zero bilinear condensate \cite{Cohen:1983fd}. However
periodic boundary conditions, at least in the continuum, only allow for smooth gauge fields of integer charge. 
Since in general the path integral has contributions from non-smooth gauge fields as well, the configurations to which the
adjoint index assigns a fractional charge at finite lattice spacing would have been good candidates for explaining the bilinear condensate
in a periodic box.

One consequence of our finding is that this is not the case. The bilinear gaugino condensate
cannot be attributed to configurations with fractional topological charge in a periodic box
because such configurations do not survive the continuum limit. 

\section{Topological susceptibility}

The topological susceptibility, $\chi = \langle Q^2 \rangle / V$, can straightforwardly be calculated from 
the distribution of indices and the corresponding charges. 
The continuum expectation is certainly that the representation used to define the topological charge is irrelevant and any two
representations should lead to the same topological susceptibility.
The fundamental index always leads to
integer charges even at finite lattice spacing, but the sextet index gives fractional charges whenever it is not a multiple of
$5$. In any case, the charge will be defined as $Q = I / 2T$ for both representations.

Table \ref{susc} shows the measured susceptibilities for both representations. They are consistent with \cite{Kovacs:2002nz} at comparable
lattice spacings. Disregarding the largest lattice spacing a linear
fit in $a^2$ gives $r_0^4 \chi_F = 0.050(8)$ and $r_0^4 \chi_S = 0.054(6)$ for the continuum values. Clearly, they are compatible
with each other and also with the continuum extrapolated value in physical volume $L = 1.49\,fm$ in \cite{Giusti:2003gf} and the
combined infinite volume and continuum extrapolated value in \cite{Durr:2006ky}. 
Our volume is $L=1.2215\,fm$ but the volume dependence is expected to be very small as long as
$L$ is larger than about $1\,fm$.

We have seen that in the sextet representation configurations where $I/2T$ is not an integer are lattice artifacts. Consequently
one might define the topological susceptibility via restricting the vacuum expectation value $\langle Q^2 \rangle / V$ to those
configurations only where $I/2T$ is an integer. This would reduce the statistics and increase errors. Nevertheless the topological
susceptibility could be
continuum extrapolated in this way as well, omitting the largest lattice spacing similarly to above. The extrapolated topological
susceptibility is $r_0^4 \chi_S^\prime = 0.048(3)$ (although with a very small $\chi^2/{\rm dof}$) which, within errors, agrees 
with the one obtained above, involving all configurations.

\begin{table}
\begin{center}
\begin{tabular}{|c|ccccc|}
\hline
$a/r_0$ & 0.2036  & 0.1745 & 0.1527 & 0.1222 & 0.1018 \\
\hline
\hline
$r_0^4\chi_F$  & 0.062(4)  & 0.066(5) & 0.066(3) & 0.059(7) & 0.049(11) \\
\hline
$r_0^4\chi_S$  & 0.060(5)  & 0.061(5) & 0.062(3) & 0.058(7) & 0.050(12) \\
\hline
$r_0^4\chi_S^\prime$  & 0.052(7) & 0.060(8) & 0.059(6)  & 0.055(7)  & 0.050(13)  \\
\hline
\end{tabular}
\caption{Topological susceptibility from both the fundamental and sextet representations. In the last row for the sextet
representation only configurations with integer $Q = I_S/5$ were considered.\label{susc}}
\end{center}
\end{table}

\section{Discussion}

Strong numerical evidence has been presented that the sextet and fundamental representation give the same integral 
definition of the topological charge in the continuum limit in a large periodic box.
What is known rigorously is that the purely geometric definition of the charge gives an
integer once a smoothness condition on the plaquettes, $||1-U_p||<\varepsilon$, is satisfied for some $\varepsilon$. 
This is a sufficient but not necessary condition \cite{Luscher:1981zq}. 

Although the plaquette condition precludes the existence of a
positive transfer matrix at finite lattice spacing \cite{Creutz:2004ir}, it also enters the rigorous treatment of the overlap operator. Its locality can
be established rigorously if this condition is imposed on all plaquettes \cite{Hernandez:1998et}. However
for locality the link variables need to be much smoother than for having an integer geometric charge, i.e.
the $\varepsilon$ needed for locality is much smaller than the one needed for an integer geometric charge.

We conjecture that if the stronger plaquette condition is imposed on the link variables, any two representations will lead to identical
and integral topological charge via the index theorem and that this integer will agree with the geometric definition. This
stronger plaquette condition is probably only sufficient but not necessary.

One byproduct of our analysis was a result on ${\cal N} = 1$ super-Yang-Mills theory on $R^4$.
The non-zero value of the bilinear gaugino condensate is certainly due to non-perturbative effects but it is entirely possible
that it does not come from semi-classical gauge fields. For example instantons are most probably not the only contribution to the non-zero
multi-linear condensate involving $2N$ gauginos either \cite{Fuchs:1985ft, Hollowood:1999qn}. In this case instantons of course do contribute but most probably there
are other non-perturbative contributions as well. The situation seems similar for the bilinear condensate the only difference being that
no contribution can be attributed to semi-classical fields and the entire quantity comes from other non-perturbative effects, at
least in a large periodic box, or equivalently on $R^4$. 

It should be noted that of course on $S^1 \times R^3$ objects with fractional topological charge
are well known to exist \cite{Kraan:1998pm, Kraan:1998sn, Bruckmann:2003ag, Bruckmann:2004nu, Poppitz:2008hr}
and have been argued to give rise to a non-zero bilinear condensate under suitable boundary conditions \cite{Davies:1999uw}.
However here we have considered the case of periodic boundary conditions which is approriate for $R^4$.

\section*{Acknowledgements}

DN would like to acknowledge useful discussions with David Adams, Tom DeGrand, Urs Heller and Ken Intriligator.
The computations were carried out on the USQCD clusters at Fermilab and the GPU clusters at Wuppertal University.
This work is supported by the NSF under
grant 0704171, by the DOE under grants DOE-FG03-97ER40546, DOE-FG-02-97ER25308,
by the DFG under grant FO 502/1 and by SFB-TR/55.

\newpage

\begin{table}
\begin{tabular}{|c|c|c|c|c|}
\hline
$I^3_F$& $I^3_S$ & $5I^3_F-I^3_S$ & $I^3_S-I^3_F$ & \#configs \\ 
\hline
\hline
1&	5&	0&	4&	52\\
\hline
0&	2&	-2&	2&	35\\
\hline
1&	3&	2&	2&	32\\
\hline
0&	0&	0&	0&	30\\
\hline
1&	7&	-2&	6&	29\\
\hline
2&	10&	0&	8&	26\\
\hline
2&	8&	2&	6&	21\\
\hline
1&	4&	1&	3&	18\\
\hline
0&	1&	-1&	1&	18\\
\hline
1&	6&	-1&	5&	17\\
\hline
2&	12&	-2&	10&	12\\
\hline
2&	11&	-1&	9&	12\\
\hline
1&	1&	4&	0&	12\\
\hline
1&	2&	3&	1&	11\\
\hline
0&	3&	-3&	3&	10\\
\hline
2&	9&	1&	7&	8\\
\hline
2&	7&	3&	5&	8\\
\hline
2&	6&	4&	4&	8\\
\hline
1&	9&	-4&	8&	8\\
\hline
0&	4&	-4&	4&	8\\
\hline
3&	13&	2&	10&	6\\
\hline
0&	5&	-5&	5&	6\\
\hline
2&	14&	-4&	12&	5\\
\hline
1&	8&	-3&	7&	5\\
\hline
3&	16&	-1&	13&	4\\
\hline
\end{tabular}
\begin{tabular}{|c|c|c|c|c|c|}
\hline
$I^3_F$& $I^3_S$ & $5I^3_F-I^3_S$ & $I^3_S-I^3_F$ & \#configs \\ 
\hline
\hline
1&	0&	5&	-1&	4\\
\hline
0&	6&	-6&	6&	4\\
\hline
3&	15&	0&	12&	3\\
\hline
2&	5&	5&	3&	3\\
\hline
4&	18&	2&	14&	2\\
\hline
4&	17&	3&	13&	2\\
\hline
3&	19&	-4&	16&	2\\
\hline
3&	17&	-2&	14&	2\\
\hline
3&	12&	3&	9&	2\\
\hline
3&	11&	4&	8&	2\\
\hline
4&	20&	0&	16&	1\\
\hline
4&	19&	1&	15&	1\\
\hline
4&	16&	4&	12&	1\\
\hline
4&	15&	5&	11&	1\\
\hline
3&	20&	-5&	17&	1\\
\hline
3&	18&	-3&	15&	1\\
\hline
3&	14&	1&	11&	1\\
\hline
3&	10&	5&	7&	1\\
\hline
3&	7&	8&	4&	1\\
\hline
2&	13&	-3&	11&	1\\
\hline
1&	12&	-7&	11&	1\\
\hline
1&	10&	-5&	9&	1\\
\hline
0&	9&	-9&	9&	1\\
\hline
0&	7&	-7&	7&	1\\
\hline
 & & & & \\
\hline
\end{tabular}
\caption{The sextet and fundamental indices for $L/a = 12$.\label{indices12}}
\end{table}

\begin{table}
\begin{tabular}{|c|c|c|c|c|}
\hline
$I^3_F$& $I^3_S$ & $5I^3_F-I^3_S$ & $I^3_S-I^3_F$ & \#configs \\ 
\hline
\hline
1&	5&	0&	4&	95\\
\hline
0&	0&	0&	0&	66\\
\hline
0&	2&	-2&	2&	49\\
\hline
2&	10&	0&	8&	48\\
\hline
1&	3&	2&	2&	37\\
\hline
2&	8&	2&	6&	27\\
\hline
1&	7&	-2&	6&	24\\
\hline
0&	1&	-1&	1&	21\\
\hline
1&	4&	1&	3&	18\\
\hline
3&	15&	0&	12&	17\\
\hline
1&	1&	4&	0&	12\\
\hline
3&	13&	2&	10&	11\\
\hline
1&	6&	-1&	5&	11\\
\hline
0&	3&	-3&	3&	11\\
\hline
2&	12&	-2&	10&	9\\
\hline
2&	9&	1&	7&	9\\
\hline
2&	6&	4&	4&	8\\
\hline
0&	4&	-4&	4&	8\\
\hline
1&	8&	-3&	7&	7\\
\hline
3&	14&	1&	11&	6\\
\hline
2&	7&	3&	5&	6\\
\hline
3&	17&	-2&	14&	5\\
\hline
2&	11&	-1&	9&	5\\
\hline
1&	2&	3&	1&	4\\
\hline
4&	20&	0&	16&	3\\
\hline
3&	16&	-1&	13&	3\\
\hline
2&	13&	-3&	11&	3\\
\hline
4&	22&	-2&	18&	2\\
\hline
4&	16&	4&	12&	2\\
\hline
3&	12&	3&	9&	2\\
\hline
3&	11&	4&	8&	2\\
\hline
2&	14&	-4&	12&	2\\
\hline
2&	0&	10&	-2&	2\\
\hline
5&	20&	5&	15&	1\\
\hline
4&	18&	2&	14&	1\\
\hline
3&	18&	-3&	15&	1\\
\hline
2&	5&	5&	3&	1\\
\hline
2&	4&	6&	2&	1\\
\hline
1&	9&	-4&	8&	1\\
\hline
1&	0&	5&	-1&	1\\
\hline
0&	5&	-5&	5&	1\\
\hline
\end{tabular}
\begin{tabular}{|c|c|c|c|c|}
\hline
$I^3_F$& $I^3_S$ & $5I^3_F-I^3_S$ & $I^3_S-I^3_F$ & \#configs \\ 
\hline
\hline
1&	5&	0&	4&	94\\
\hline
0&	0&	0&	0&	58\\
\hline
2&	10&	0&	8&	54\\
\hline
1&	3&	2&	2&	30\\
\hline
1&	7&	-2&	6&	19\\
\hline
0&	2&	-2&	2&	19\\
\hline
1&	4&	1&	3&	15\\
\hline
3&	13&	2&	10&	14\\
\hline
2&	8&	2&	6&	14\\
\hline
0&	1&	-1&	1&	14\\
\hline
3&	15&	0&	12&	12\\
\hline
2&	9&	1&	7&	8\\
\hline
1&	6&	-1&	5&	6\\
\hline
0&	4&	-4&	4&	6\\
\hline
3&	17&	-2&	14&	4\\
\hline
2&	12&	-2&	10&	4\\
\hline
2&	11&	-1&	9&	4\\
\hline
0&	3&	-3&	3&	4\\
\hline
4&	20&	0&	16&	3\\
\hline
2&	6&	4&	4&	3\\
\hline
1&	9&	-4&	8&	3\\
\hline
1&	1&	4&	0&	3\\
\hline
4&	16&	4&	12&	2\\
\hline
3&	14&	1&	11&	2\\
\hline
2&	5&	5&	3&	2\\
\hline
2&	4&	6&	2&	2\\
\hline
4&	22&	-2&	18&	1\\
\hline
4&	18&	2&	14&	1\\
\hline
3&	18&	-3&	15&	1\\
\hline
3&	12&	3&	9&	1\\
\hline
2&	7&	3&	5&	1\\
\hline
1&	10&	-5&	9&	1\\
\hline
1&	8&	-3&	7&	1\\
\hline
1&	2&	3&	1&	1\\
\hline
1&	0&	5&	-1&	1\\
\hline
 & & & & \\
\hline
 & & & & \\
\hline
 & & & & \\
\hline
 & & & & \\
\hline
 & & & & \\
\hline
 & & & & \\
\hline
\end{tabular}
\caption{The sextet and fundamental indices for $L/a = 14$ (left) and $L/a = 16$ (right).\label{indices1416}}
\end{table}

\begin{table}
\begin{center}
\begin{tabular}{|c|c|c|c|c|}
\hline
$I^3_F$& $I^3_S$ & $5I^3_F-I^3_S$ & $I^3_S-I^3_F$ & \#configs \\ 
\hline
\hline
1&	5&	0&	4&	64\\
\hline
0&	0&	0&	0&	51\\
\hline
2&	10&	0&	8&	32\\
\hline
3&	15&	0&	12&	14\\
\hline
1&	7&	-2&	6&	10\\
\hline
0&	2&	-2&	2&	6\\
\hline
2&	8&	2&	6&	5\\
\hline
1&	4&	1&	3&	5\\
\hline
1&	3&	2&	2&	5\\
\hline
0&	1&	-1&	1&	5\\
\hline
2&	12&	-2&	10&	4\\
\hline
3&	13&	2&	10&	3\\
\hline
4&	18&	2&	14&	1\\
\hline
3&	17&	-2&	14&	1\\
\hline
3&	14&	1&	11&	1\\
\hline
2&	9&	1&	7&	1\\
\hline
1&	6&	-1&	5&	1\\
\hline
1&	2&	3&	1&	1\\
\hline
1&	1&	4&	0&	1\\
\hline
\end{tabular}
\begin{tabular}{|c|c|c|c|c|}
\hline
$I^3_F$& $I^3_S$ & $5I^3_F-I^3_S$ & $I^3_S-I^3_F$ & \#configs \\ 
\hline
\hline
1	 & 5	 & 0	 & 4	 & 41 \\
\hline
0	 & 0	 & 0	 & 0	 & 28 \\
\hline
2	 & 10	 & 0	 & 8	 & 14 \\
\hline
3	 & 15	 & 0	 & 12	 & 3 \\
\hline
2	 & 12	 & -2	 & 10	 & 3 \\
\hline
1	 & 3	 & 2	 & 2	 & 3 \\
\hline
4	 & 20	 & 0	 & 16	 & 2 \\
\hline
1	 & 2	 & 3	 & 1	 & 2 \\
\hline
1	 & 7	 & -2	 & 6	 & 2 \\
\hline
0	 & 1	 & -1	 & 1	 & 1 \\
\hline
0	 & 2	 & -2	 & 2	 & 1 \\
\hline
0	 & 3	 & -3	 & 3	 & 1 \\
\hline
1	 & 4	 & 1	 & 3	 & 1 \\
\hline
1	 & 6	 & -1	 & 5	 & 1 \\
\hline
2	 & 8	 & 2	 & 6	 & 1 \\
\hline
2	 & 9	 & 1	 & 7	 & 1 \\
\hline
 & & & & \\
\hline
 & & & & \\
\hline
 & & & & \\
\hline
\end{tabular}
\caption{The sextet and fundamental indices for $L/a = 20$ (left) and $L/a = 24$ (right).\label{indices20}}
\end{center}
\end{table}

\end{document}